\documentclass[aps,prl,twocolumn,showpacs,superscriptaddress]{revtex4-1}
\usepackage{graphicx} 
\usepackage{dcolumn}   
\usepackage{bm}      
\usepackage{amssymb}   
\usepackage{ulem}

\begin{document}
\title{Dynamic synchrotron X-ray imaging study of effective temperature in a vibrated granular medium}
\author{Yixin Cao}
\affiliation{Department of Physics and Astronomy, Shanghai Jiao Tong University - 800 Dongchuan Rd. Shanghai 200240, China}
\author{Xiaodan Zhang}
\affiliation{Department of Physics and Astronomy, Shanghai Jiao Tong University - 800 Dongchuan Rd. Shanghai 200240, China}
\author{Binquan Kou}
\affiliation{Department of Physics and Astronomy, Shanghai Jiao Tong University - 800 Dongchuan Rd. Shanghai 200240, China}
\author{Xiangting Li}
\affiliation{Department of Physics and Astronomy, Shanghai Jiao Tong University - 800 Dongchuan Rd. Shanghai 200240, China}
\author{Xianghui Xiao}
\affiliation{X-ray Science Division, Argonne National Laboratory, 9700 South Cass Avenue, IL 60439, USA}
\author{Kamel Fezzaa}
\affiliation{X-ray Science Division, Argonne National Laboratory, 9700 South Cass Avenue, IL 60439, USA}
\author{Yujie Wang}
\email{yujiewang@sjtu.edu.cn}
\affiliation{Department of Physics and Astronomy, Shanghai Jiao Tong University - 800 Dongchuan Rd. Shanghai 200240, China}
\date{\today}
\begin{abstract}
We present a dynamic synchrotron X-ray imaging study of the effective temperature $T_{eff}$ in a vibrated granular medium. By tracking the directed motion and the fluctuation dynamics of the tracers inside, we obtained $T_{eff}$ of the system using Einstein relation. We found that as the system unjams with increasing vibration intensities $\Gamma$, the structural relaxation time $\tau$ increases substantially which can be fitted by an Arrhenius law using $T_{eff}$. And the characteristic energy scale of structural relaxation yielded by the Arrhenius fitting is $E = 0.21 \pm 0.02$ $pd^3$, where $p$ is the pressure and $d$ is the background particle diameter, which is consistent with those from hard sphere simulations in which the structural relaxation happens via the opening up of free volume against pressure.
\end{abstract}
\date{\today}
\maketitle
The effective temperature $T_{eff}$ has attracted a lot of interest in the study of out-of-equilibrium glassy systems like structural glass, colloids, foams, and granular materials\cite{cug}. The introduction of $T_{eff}$ can help the understanding of the aging and various transport phenomena\cite{cug} in these systems. This is evidenced by the fact that the structural relaxation processes under shear in these out-of equilibrium systems are controlled by the long-time scale $T_{eff}$ instead of the short-time scale kinetic temperature $T_{k}$\cite{ilg,cug2,barrat_chem,ohern}. Additionally, the study of $T_{eff}$ can lead to a unified understanding of the jamming phase diagram\cite{jamshear} and the plastic deformation of solids under shear\cite{tom_liu,tom}. The usefulness of $T_{eff}$ has also been validated by the fact that various definitions have yielded consistent values which makes it easy for experimental measurements\cite{barrat_pre,ono,makse_nt}. Recently, $T_{eff}$ has become one of the key concepts in the development of mesoscopic thermodynamic theories of amorphous solid plasticity and soft glassy rheology\cite{langer_rev,sollich_pre}.

Granular systems are by nature out-of-equilibrium systems since they will simply come to rest without outside energy input. However, when agitated by shaking or shear, they can display gas-, fluid- and solid-like phases under different energy input strength, which prompts possible thermodynamic description of these phases. Kinetic theory originally based on ideal gas has been quite successful in describing highly agitated dilute granular gases after taking into account the dissipation\cite{jenkins}. In the dilute limit, $T_{eff}$ based on the fluctuation-dissipation theorem can be defined, which justifies a thermodynamic approach\cite{danna} despite energy non-equipartition and velocity non-Gaussian distribution\cite{menon}. In this case, $T_{eff} \propto \Gamma^2$ where $\Gamma$ is the vibration intensity and $T_{eff}$ is on the same order of magnitude as the particle's mean kinetic energy\cite{danna}. At the other limit, when the outside energy input is absent, the granular system undergoes a jamming phase transition into a disordered solid phase\cite{jamming}. Edwards first suggested that when a granular packing is slowly sheared, the system explores the stable mechanical states which is the same as taking a flat average of the jammed configurations\cite{edwards1989}. The corresponding configurational temperature defined based on this ensemble turns out to be equivalent to the temperature $T_{eff}$ based on fluctuation-dissipation theorem\cite{makse_nt}. In practice, volume or stress instead of energy is normally considered as the conserved quantity and concepts similar to temperature like compactivity, angoricity or a combination of these two have been introduced and studied\cite{makse_nt,edwards1989,edwards2005,edwards2009,henkes,edwards2012,makse2012}. Experimentally, this thermodynamic approach has been adopted in granular compaction studies in which the vibration intensity $\Gamma$ has been interpreted as the temperature-like parameter\cite{nmat} similar to the granular gas case\cite{danna}. Between the slowly sheared state and the highly-agitated granular gas, the system is in a dense liquid state in which both particle collisions and perpetual contacts are important. This regime is important in many scientific and industrial applications and empirical constitutive relations have been introduced to describe its rheological behaviour\cite{midi,jop}. However, a more fundamental theory based on microscopic dynamics is needed to justify these models. Therefore, it is interesting to see whether a thermodynamic theory is still valid to the different phases of a granular system when it is sheared or shaken with increasing strength as it evolves from a static packing into a granular gas, $i.e.$, whether a valid and consistent $T_{eff}$ can be defined for all phases. It is also important to understand what determines $T_{eff}$ and how does it influence the glassy transport properties and solid plasticity\cite{tom_liu,langer_rev,langer_pre,sollich_prl}.

In the current study, we investigated $T_{eff}$ inside a three-dimensional (3D) mechanically driven granular system when it evolves from a dense granular fluid to a granular gas. By introducing tracers different from the background particles, we can monitor their trajectories inside the 3D granular medium non-invasively and dynamically using synchrotron X-ray imaging technique\cite{yjwang,fezzaa}. We can observe the otherwise invisible particle motions inside the granular medium with high spatial and temporal resolutions. We obtained $T_{eff}$ of the system using Einstein relation by tracking the directed motion and the fluctuation dynamics of the tracers inside. It has been observed that as the system unjams, the structural relaxation time $\tau$ increases substantially and an Arrhenius fitting of $\tau$ versus $T_{eff}$ yields a characteristic activation dynamics energy scale similar to those from hard sphere simulations. It suggests that the structural relaxation in dense granular fluid can be interpreted similar to those in hard spheres as opening up free volume against background pressure\cite{tom,tom_liu,xu}.

\section{Experimental setup and imaging technique}
\begin{figure}[h]
\centering
  \includegraphics[width=0.5\textwidth,trim=0cm 0cm 0cm 0cm]{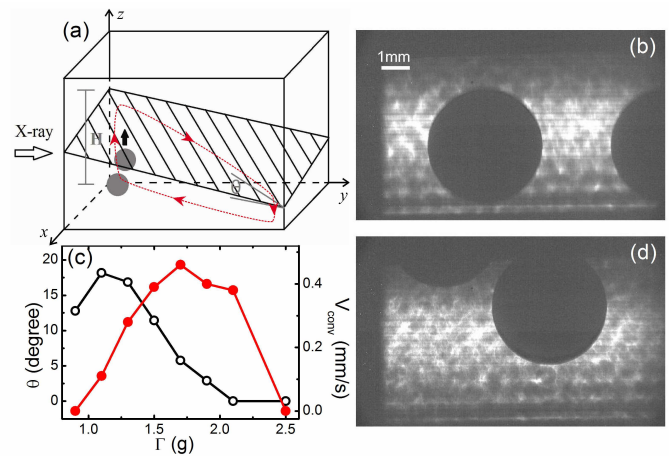}
  \caption{(Color online) (a) Schematic of the experimental setup. The granular medium in the container develops a strong surface tilting and convection. Two intruder particles are placed near the corner which has the largest surface height $H$. The surface tilting is characterized by the surface slope angle $\theta$ and the convection direction is indicated by the red dashed line. (b) and (d) show X-ray images of the tracers inside the acrylic container filled with glass particles before and after extended period of vibration. (c) shows surface slope angle $\theta$ (open circles) and the convection speed $V_{conv}$ (solid circles) as functions of vibration intensity $\Gamma$.}
  \label{fig.1}
\end{figure}

The experiment was carried out at 2BM beam line of the Advanced Photon Source of Argonne National Laboratory. The brilliant unfiltered ``pink'' X-ray beam from the synchrotron ring is utilized for the dynamic X-ray studies. The imaging system consists of a fast LAG scintillator coupled to a high-speed Cooke Dimax CMOS camera ($11-\mu$m pixel, $2016 \times 1216$ pixel array) via a $2 \times $ microscope objective. The effective X-ray field-of-view is $11 \times 6$ $\textrm{mm}^2$. The imaging system was placed $0.35$ m away from the sample to optimize the phase-contrast effects, which are very useful in the detection of the boundaries of the granular particles. A smooth acrylic container which has a square base with area $A_{sys}=28 \times 28$ $\textrm{mm}^2$ was filled up to a height of $H=10$ mm using polydisperse glass particles with diameter $d = 0.73 \pm 0.17$ mm and density $\rho = 2.7$ g/cm$^3$ (see schematic in Fig.~\ref{fig.1}(a)). Two steel tracer balls with diameter $d_{tr}=4.1$ mm and density $\rho_{tr} = 7.9$ g/cm$^3$ were buried near one corner of the container bottom. Subsequently, the container was mounted on an electromagnetic exciter which vibrates sinusoidally at a fixed frequency $f=50$ Hz. The effective vibration intensity $\Gamma = A(2\pi f)^2$ was varied from $0.9$ g to $2.5$ g by changing the vibration amplitude $A$, where g is the gravitational acceleration.

In Fig.~\ref{fig.1}(b) and (d), X-ray images of the granular medium (tracers within the X-ray field-of-view) before and after more than thirty-minutes of vibration are shown. The small glass particles appear as a speckle background in the images due to the X-ray multiple-scattering effects. The distinctive contrast between the tracers and the background is owing to the large X-ray absorption coefficient difference between steel and glass, which greatly facilitates the identification of the locations and speeds of the tracers.

\section{Convection and surface tilting}
Macroscopic convection roll and surface tilting are both present in the system with the former serving as the major mechanism of Brazil nut effect for the ascension movements of the tracers under shaking\cite{kudrolli,swinney}. As the vibration intensity is gradually increased, the granular medium first develops a large surface tilting angle $\theta$ at $\Gamma=0.9$ g without macroscopic flow (see fig.~\ref{fig.1}(c)). The granular bed is not fluidized and two tracers remain trapped at their original positions close to the container bottom. As $\Gamma$ increases to $1.1$ g, $\theta$ gradually decreases and a single convection roll develops, which drags the tracers to move upward. This is consistent with previous study that the unjamming transition is concurrent with the appearance of convection and surface tilting\cite{bideau}. The convection speeds $V_{conv}$ are calculated by averaging the directed motion of background glass particles within the convection roll whose trajectories could be tracked at different $\Gamma$. As shown in Fig.~\ref{fig.1}(c), the convection speed reaches maximum at intermediate $\Gamma$. When $\Gamma$ is above $2.5$ g, the granular medium turns into a granular gas and both surface tilting and convection disappear. Visual inspection from the top reveals that only rapid colliding and rattling motions of the granular particles can be seen which is different from the convective regime where the particles are in seemingly permanent contacts with each other during the flow. The progression of the phenomena suggests that as $\Gamma$ increases, the system first unjams, then turns into a dense granular fluid, and subsequently into a granular gas.

\section{Motion characteristics of the tracers}
We took X-ray images at an imaging speed of $150$ fps and tracked the tracers' displacements along both $x$ and $z$ directions (see Fig.~\ref{fig.1}(a)) using an image processing routine. To avoid the artefacts brought by tracking the particle displacements at different phases of the vibration, we only analyse the images at the same phases which yields an effective imaging speed of $50$ fps. Tracers' motions under three typical $\Gamma$ are shown in the movies of the supplementary materials. We also checked the reproducibility of the vibration motion by monitoring objects fixed on the shaker and found its position variance from different vibration cycles are negligible.

\begin{figure}[h]
\centering
  \includegraphics[scale=0.32,trim=0cm 0cm 0cm 0cm]{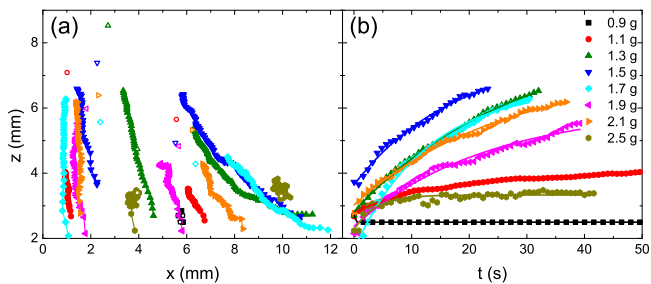}
\caption{(Color online) (a) The tracers' trajectories at different $\Gamma$. Open symbols denote relative final position after thirty-minutes of vibration. (b) The height $z$ of the center of mass of the left tracer at different $\Gamma$ as a function of time $t$, and the corresponding fitting by an exponential law.}
\label{fig.2}
\end{figure}

\begin{figure}[h]
\centering
  \includegraphics[scale=0.32,trim=0cm 0cm 0cm 0cm]{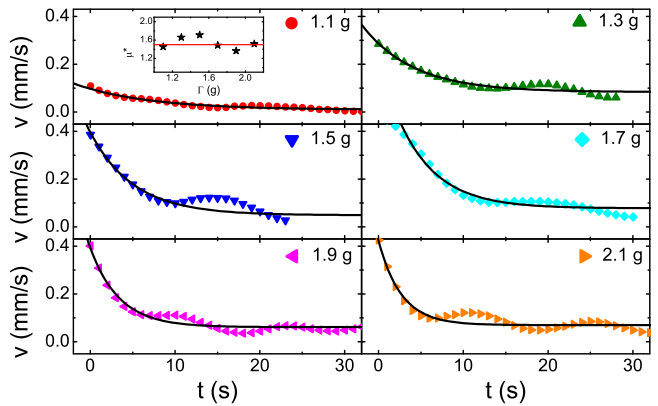}
\caption{(Color online) Left tracer's velocity along $z$-direction versus time. Solid lines are exponential fits according to Eq.~(\ref{eq.vt}). Inset shows friction coefficient $\mu^*$ calculated at different $\Gamma$ using the left tracer. Solid line indicates the mean value.}
\label{fig.3}
\end{figure}

Figure ~\ref{fig.2}(a) shows the trajectories of both tracers in the $x-z$ plane. Due to the initial position difference in the convection roll, the left tracer shows an almost $z$-direction motion while the right one has a large $x$-direction displacement component. The terminal equilibrium positions of the tracers were recorded after more than thirty-minutes of vibration and are marked by open symbols in Fig.~\ref{fig.2}(a). Fig.~\ref{fig.2}(b) plots the left tracer's vertical height $z$ vs. time $t$ for different $\Gamma$. Similar to the observation in a split-bottom Couette shear experiment, the tracers' vertical trajectory $z(t)$ seems to satisfy an exponential behavior\cite{hecke_prl,hecke_pre} which suggests an viscous-type of force is in action. By numerically differentiating the trajectory curves, we obtain the $v-t$ curves as shown in Fig.~\ref{fig.3}, where $v=\mathrm{d}z/\mathrm{d}t$ is the tracer's vertical velocity.

The curves can be fitted by an exponential law
\begin{equation}
\label{eq.vt}
v=(v_i-v_f)\exp(-t/\tau_0)+v_f,
\end{equation}
where $v_i$ ($v_f$) is the initial (final) velocity, and $\tau_0$ is a characteristic time.

\section{Equation of motion and effective viscosity}
In order to obtain $T_{eff}$ using the Einstein relation $T_{eff}=3\pi \eta d_{tr}D/k_{B}$, where $\eta$ is the viscosity and $D$ is the diffusion constant\cite{bob,zik}, we obtain $\eta$ and $D$ by measuring the viscous drag force on the tracer and its fluctuation dynamics respectively. There is still no generally accepted drag force model in granular medium for both fluidized or nonfluidized granular systems despite long-term study. Empirical constitutive relation using dynamic friction coefficient based on inertial number has been quite successful in describing the dense flow regime\cite{midi}. Recent study has also found that the effective friction coefficient increases with the drag velocity in addition to a linear depth dependency\cite{clement}. Similarly, it has been observed that a granular system under shear behaves very similarly to a simple fluid which satisfies typical Archimedes' law and has well-defined effective viscosity\cite{hecke_prl,hecke_pre,anurag}. In the following, we adopt a force model to account for the aforementioned exponential law observed. We separate the drag force into a frictional term and a ``viscous'' term\cite{kinetic}. We assume that there are three forces in action: namely, the effective gravity of tracer $M^*$g, the Coulomb friction force $F_c$, and the viscous drag force $F_{\eta}$ due to convection. The tracers' movements are over-damped, so the inertial effects can be neglected. $i.e.$, the sum of the Coulomb friction force and viscous drag force will balance the tracer's effective gravity,
\begin{equation}
\label{eq.1}
F_c+F_\eta-M^*g=0.
\end{equation}

It is well-known that the Coulomb friction force $F_c$ has a linear depth-dependency\cite{clement} with the form $F_c=\mu^* \frac{\pi}{4} d_{tr}^2\rho_{g}g(z_{surf}-z)$, where $\mu^*$ is the effective friction coefficient, here we assume $\mu^*$ is a constant without any drag velocity dependency, $\rho_{g}$ is the effective density of the granular medium, $z_{surf}$ is the surface height, and $z=z(t)$ is the tracer's height at time $t$. We also adopt a Stokes-type viscous drag force $F_{\eta}=\eta d_{tr}(V_{conv}-\frac{\mathrm{d}z}{\mathrm{d}t})$. We insert the expressions of $F_c$ and $F_{\eta}$ into Eq.~(\ref{eq.1}) and simplify it to have the form 

\begin{equation}
\label{eq.2}
z=-C_1\eta\frac{\mathrm{d}z}{\mathrm{d}t}+C_2,
\end{equation}
where $C_1=4/(\mu^* \pi d_{tr}\rho_{g}g)$ and $C_2=z_{surf}+4(\eta d_{tr}V_{conv}-M^*g)/(\mu ^* \pi d_{tr}^2\rho_{g}g)$ are constants, $M^*g=(\rho_{tr}-\rho_g)g\pi d_{tr}^3/6$.

It is obvious there exists two unknown parameters in the equation, including both $\mu^*$ and $\eta$. However, the magnitude of these two parameters can be determined since they are proportional to each other as expressed in Eq.~\ref{eq.4} when we try to match the time constant of Eq.~\ref{eq.2} with the experimentally measured $\tau_0$. Additionally, the two forces have to balance the tracer's gravity as expressed in Eq.~\ref{eq.5}, which specifies the force balance equation the tracer satisfied at $t=2\tau_0$. In the current study, we obtain $\mu^*$and $\eta$ by solving these two equations,
\begin{equation}
\tau_0=C_1\eta=(4\eta)/(\mu^*\pi d_{tr}\rho_gg),
\label{eq.4}
\end{equation}
{\small
\begin{equation}
\eta d_{tr}(V_{conv}-v\vert_{t=2\tau_0})=M^*g-\mu^*\frac{\pi}{4}d_{tr}^2\rho_gg(z_{surf}-z\vert_{t=2\tau_0}).
\label{eq.5}
\end{equation}
}

One thing to note is that we have assumed that both $\eta$ and $V_{conv}$ are constant within the narrow $z$ range investigated. The solution yields $\mu^*=1.5 \pm 0.2$ as shown in the inset of Fig.~\ref{fig.3}, and relative values used in calculation are listed in Table~\ref{tbl}. The rather constant value of $\mu^*$ for different $\Gamma$ suggests the consistency of our force model. Interestingly, the value of $\mu^*$ is larger than the static frictional one $0.45$ as determined by repose angle measurement which is also observed in previous measurements\cite{clement1,clement}. We notice that in our system, the tracers in most cases do not rise all the way to the top of the surface and the calculated magnitude of $\mu^*$ and $\eta$ are consistent with this observation, $e.g.$, when $\Gamma=1.7$ g, in the early stage, the frictional force is $17.9\times 10^{-4}$ N, accounting for about $80\%$ of the effective gravity while the viscous drag force is $4.3\times 10^{-4}$ N, which accounts for the rest $20\%$. As the tracer rises to equilibrium position when the drag velocity is maximum, the viscous drag force increase to $6.6\times 10^{-4}$ N. However, it still cannot balance the gravity alone.

\begin{figure}[h]
\centering
  \includegraphics[scale=0.32,trim=0cm 0cm 0cm 0cm]{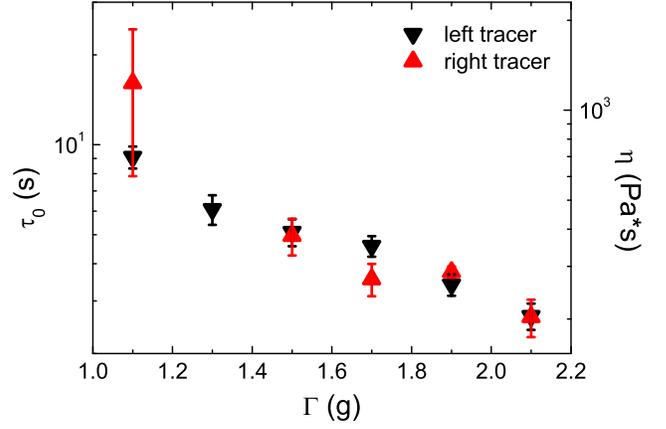}
\caption{(Color online) Characteristic time $\tau_0$ or viscosity $\eta$ versus vibration intensity $\Gamma$.}
\label{fig.4}
\end{figure}

The resulting $\tau_0$ and the corresponding $\eta$ is shown in Fig.~\ref{fig.4}. It is observed that as $\Gamma$ decreases, $\eta$ increases by about a decade.

\begin{table*}
\small
\caption{\ Main parameters of the left tracer at different vibration intensities}
\label{tbl}
	\begin{tabular*}{\textwidth}{@{\extracolsep{\fill}}cccccccccccc}
    \hline
    $\Gamma$ & $\theta$ & $V_{conv}$ & $v\vert_{t=2\tau_0}$ & $z\vert_{t=2\tau_0}$ & $\tau_0$ & $\mu^*$ & $\eta$ & $D$ & $\tau$ & $k_BT_{eff}$ & $k_BT_{k}^{low}$ \\
    (g) & ($^{\circ}$) & (mm/s) & (mm/s) & (mm) & (s) &  & (Pa$\cdot$s) & ($10^{-11}$ m$^2$/s) & (s) & ($10^{-8} J$) & ($10^{-11} J$)\\
    \hline
    0.9 & 12.8 & 0 & / & / & / & /  & / & / & / & / & / \\
    1.1 & 18.2 & 0.11 & 0.02 & 3.6 & 9.1 $\pm$ 0.8 & 1.5 & 698 $\pm$ 59 & 7.4 & 1197.8 & 0.20 $\pm$ 0.02 & 1.1\\
    1.3 & 16.9 & 0.28 & 0.11 & 4.7 & 6.1 $\pm$ 0.7 & 1.7 & 466 $\pm$ 53 & 81.5 & 109.0 & 1.5 $\pm$ 0.2 & 2.1\\
    1.5 & 11.4 & 0.39 & 0.09 & 5.4 & 5.1 $\pm$ 0.5 & 1.7 & 391 $\pm$ 41 & 122.5 & 72.5 & 1.9 $\pm$ 0.2 & 7.6\\
    1.7 & 5.8 & 0.46 & 0.16 & 4.3 & 4.6 $\pm$ 0.4 & 1.5 & 351 $\pm$ 28 & 235.0 & 37.8 & 3.2 $\pm$ 0.3 & 10.8\\
    1.9 & 2.9 & 0.40 & 0.11 & 3.3 & 3.4 $\pm$ 0.3 & 1.4 & 260 $\pm$ 21 & 229.0 & 38.8 & 2.3 $\pm$ 0.2 & 15.1\\
    2.1 & 0 & 0.38 & 0.11 & 3.8 & 2.7 $\pm$ 0.3 & 1.5 & 205 $\pm$ 21 & 233.0 & 38.1 & 1.9 $\pm$ 0.2 & 12.4\\
    2.5 & 0 & 0 & / & / & / & /  & / & / & / & / & 493.0\\
    \hline
    \end{tabular*}
\end{table*}

\section{Fluctuation and diffusion dynamics}
\begin{figure}[h]
\centering
  \includegraphics[scale=0.32,trim=0cm 0cm 0cm 0cm]{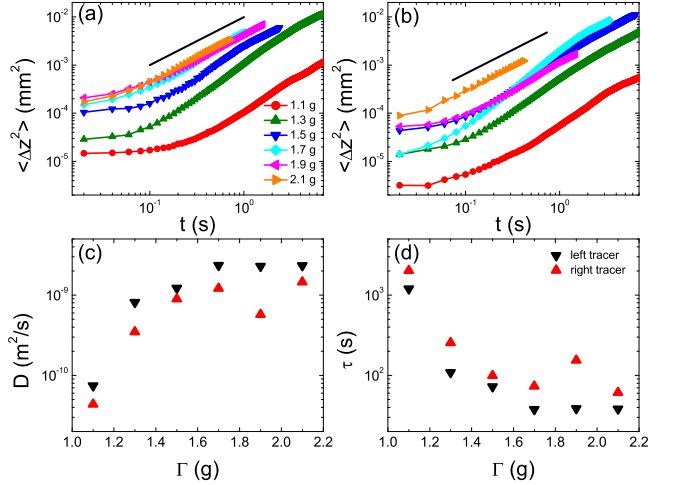}
\caption{(Color online) Diffusion dynamics of (a) the left and (b) the right tracers along $z$-direction. Different symbols represent different vibration intensities. The black solid line has slope of one. (c) Diffusion constant as a function of vibration intensity. (d) Structural relaxation time as a function of vibration intensity.}
\label{fig.5}
\end{figure}

We studied the tracers' diffusion dynamics along $z$-direction. To obtain the fluctuating dynamics only, we subtract trajectories along $z$-direction by their smoothed counterparts using exponential fits. The corresponding mean square displacement (MSD) as a function of $t$ under different $\Gamma$ is shown in Fig.~\ref{fig.5}(a) and (b).

The diffusive dynamics is clearly established and from each curve we can extract a diffusion constant $D$. We also define the structural relaxation time $\tau$ as the corresponding time scale when the MSD of each curve equals $\frac{1}{3}d^2$. These two parameters are plotted in Fig.~\ref{fig.5} (c) and (d) respectively as functions of $\Gamma$. Overall, the diffusion constants $D$ increases as $\Gamma$ increases initially and saturates at large $\Gamma$, while $\tau$ has a decreasing trend as $\Gamma$ increases and also saturates at large $\Gamma$.

\section{Effective temperature and activation energy}
\begin{figure}[h]
\centering
  \includegraphics[scale=0.32,trim=0cm 0cm 0cm 0cm]{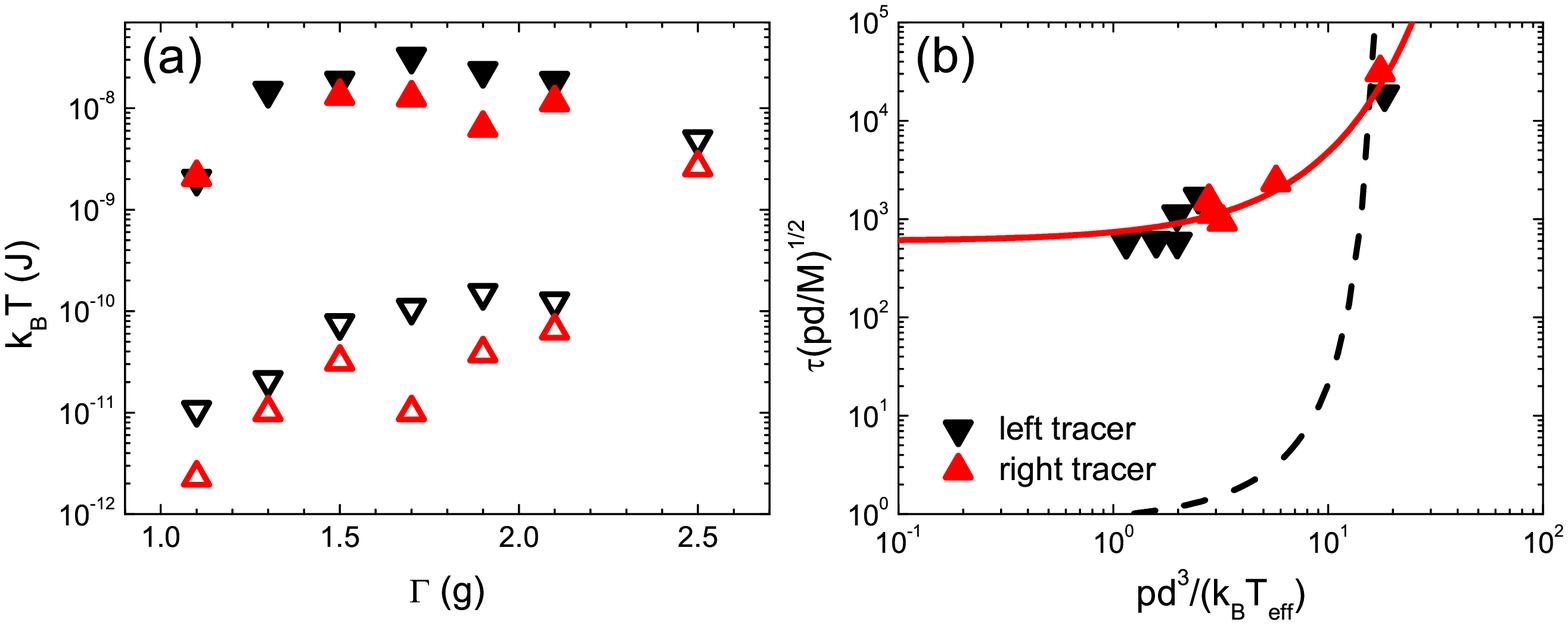}
\caption{(Color online) (a) Effective temperature $T_{eff}$ (solid) and a lower bound kinetic temperature $T_{k}^{low}$ (open) versus vibration intensity $\Gamma$. (b) the normalized structural relaxation time as a function of normalized inverse effective temperature. Solid line corresponds to Arrhenius fit as Eq.(\ref{eq.3}) and dashed line corresponds to Vogel-Fulcher fit of equilibrium data adopted from \cite{tom,xu}.}
\label{fig.6}
\end{figure}

The measured $k_{B}T_{eff}$ remains above $10^{-9} J$ for the wide range of $\Gamma$ studied as shown in fig.~\ref{fig.6}(a) and listed in Table~\ref{tbl}. This energy roughly equals $mgd \simeq 4 \times 10^{-9} J$ where $m$ is the glass particle mass, which is reminiscent of the measurement using a Couette cell under constant shear\cite{makse2005}. To understand the significance of $T_{eff}$, we studied how structural relaxation time $\tau$ is dependent on $T_{eff}$. Similar approach has been adopted to study the compaction process of granular packings under tapping\cite{nmat}, where $\Gamma$ is used instead of $T_{eff}$ and an Arrhenius law has been adopted\cite{reddy}. In the following, we follow the same strategy and compare our results with those from a hard sphere simulation on a more quantitative fashion. To be consistent with the dimensionless results from the hard sphere simulation, we normalize $T_{eff}$ and $\tau$ with the typical energy scale $pd^3$ and time scale $\sqrt{pd/M}$. We notice that our system can be assumed to be under fairly constant pressure $p=Nmg/A_{sys} \simeq 95$ Pa when the tracer position is around $z=4$ mm, and $N=\phi A_{sys}(z_{surf}-z)/(\pi d^3/6)$ is the approximate number of glass particles above the tracer in the system with packing fraction $\phi \approx 0.60$.
Once $p$ is known, we obtain the dimensionless relation between $\tau$ and $T_{eff}$ as shown in Fig.~\ref{fig.6}(b). The divergence of $\tau$ towards jamming is fitted using a simple Arrhenius law
\begin{equation}
\label{eq.3}
y\propto\exp(\frac{E}{x}),
\end{equation}
where $x=k_BT_{eff}/(pd^3)$ and $y=\tau\sqrt{pd/M}$. The fitting result yields an activation dynamics energy scale $E=0.21 \pm 0.02$ $pd^3$, which we found to be consistent with that of out-of-equilibrium hard sphere fluid under shear\cite{tom} where $E=0.11\sim 0.22$ $pd^3$ and that of thermal hard sphere fluid\cite{xu} where $E=0.18\sim 0.25$ $pd^3$. This suggests that the structural relaxation energy scale in vibrated dense granular medium is similar to those in hard spheres systems where structural relaxation happens by opening up of free volume against the pressure\cite{tom,xu}.

\section{Kinetic temperature}
We also define a lower bound estimate of the kinetic temperature $k_{B}T_{k}^{low}=M\left\langle\delta v^2\right\rangle$ of the tracers using the mean square fluctuating velocity along $z$-direction during a time period of $0.02$ s. This corresponds to our shortest time resolution in resolving the particle displacements. The reason $T_k^{low}$ is a lower bound is due to the fact that even in dense flows, ballistic motions of granular particles can still be present and will lead the real $T_k$ higher than $T_k^{low}$. The measured $T_{k}^{low}$ is shown in fig.~\ref{fig.6}(a). It is clear that $T_k^{low}$ is substantially lower than $T_{eff}$ over the whole $\Gamma$ range and only reaches the same order of magnitude with our measured $T_{eff}$ when the system turns into a gas state at $\Gamma=2.5$ g. At this $\Gamma$, the dominating particle motions are colliding motions. This contrasts with the small $\Gamma$ regime when the particles are seemingly in frictional motions against each other. When $\Gamma=2.5$ g, $T_k^{low}$ is on the same order of magnitude as $T_{eff}$ which also suggests that the system reaches thermal equilibrium between its short- and long-time dynamics when it turns into a granular gas.

\section{Conclusions}
In the current study, we have examined the complex behaviour of a dense granular system under vibration very close to the jamming density. It is found that close to jamming, the system does not show a simple $T_{eff}\propto \Gamma^2$ in highly agitated granular medium\cite{danna} where the $T_{eff}$ is on the same order of magnitude as the tracers' kinetic temperature $T_k$. Instead, we observe that $T_{eff}$ is controlled by pressure $p$\cite{xu_ohern} which is owning to the possible universal structural relaxation mechanism similar to hard spheres\cite{tom,xu}. We also notice that similar energy scale could be relevant for plastic deformation. How is $T_{eff}$ related to STZ\cite{langer_rev} or SGR theory\cite{sollich_prl} remains future study.

\section{Acknowledgement}
We thank Thomas K. Haxton and Ning Xu for many helpful discussions. Some of the initial work has been carried out at BL13W1 beamline of Shanghai Synchrotron Radiation Facility (SSRF), and the work is supported by the Chinese National Science Foundation No. 11175121, and National Basic Research Program of China (973 Program; 2010CB834301). The use of the APS was supported by the U. S. Department of Energy, Office of Science, Office of Basic Energy Sciences, under Contract No. DE-AC02-06CH11357.
\bibliography{rsc}

\end{document}